\documentclass[12pt]{article}

\usepackage{epsfig,amsfonts,amssymb}
\newcommand{\beq}{\begin{equation}}
\newcommand{\eeq}{\end{equation}}
\newcommand{\be}{\begin{equation}}
\newcommand{\ee}{\end{equation}}
\newcommand{\beqa}{\begin{eqnarray}}
\newcommand{\eeqa}{\end{eqnarray}}
\newcommand{\beqar}{\begin{eqnarray*}}
\newcommand{\eeqar}{\end{eqnarray*}}
\newcommand{\eg}{{\it e.g.,}\ }
\newcommand{\ie}{{\it i.e.,}\ }
\newcommand{\labell}[1]{\label{#1}} 
\newcommand{\reef}[1]{(\ref{#1})}

\begin{document}

\thispagestyle{empty}


\hfill{hep-th/0608012}

\hfill{}

\hfill{}

\hfill{}

\hfill{}

\vspace{32pt}

\begin{center}
\textbf{\Large Black Rings}\\

\vspace{40pt}

Roberto Emparan$^{a,b}$ and
Harvey S.~Reall$^c$

\vspace{12pt}
$^a$\textit{Instituci\'o Catalana de Recerca i Estudis Avan\c cats (ICREA)}\\
\vspace{6pt}
$^b$\textit{Departament de F{\'\i}sica Fonamental}\\
\textit{Universitat de
Barcelona, Diagonal 647, E-08028, Barcelona, Spain}\\
\vspace{6pt}
$^c$\textit{School of Physics and Astronomy, University of Nottingham, NG7 2RD, UK}\\
\vspace{6pt}
\texttt{emparan@ub.edu, harvey.reall@nottingham.ac.uk}
\end{center}

\vspace{15pt}

\begin{center}
{\sl Dedicated to the memory of Andrew Chamblin}
\end{center}

\vspace{20pt}

\begin{abstract}

A black ring is a five-dimensional black hole with an event horizon of
topology $S^1\times S^2$. We provide an introduction to the description
of black rings in general relativity and string theory. Novel aspects of
the presentation include a new approach to constructing black ring
coordinates and a critical review of black ring microscopics.

\end{abstract}

\setcounter{footnote}{0}

\newpage

\tableofcontents

\newpage

\section{Introduction}

The classical theory of black holes developed in the 1960's-70's
produced a proof of the simplicity of four-dimensional black holes. An
asymptotically flat, stationary black hole solution\footnote{Asymptotic
flatness and stationarity will be assumed of all solutions considered in
this paper.} of four-dimensional Einstein-Maxwell theory is fully
specified by a handful of parameters---the black hole has `no hair'.
More strongly, the uniqueness theorems assert that these parameters are
precisely those that correspond to conserved charges, namely, the mass
$M$ and angular momentum $J$, and possibly the charges $Q$ associated to
local gauge symmetries. Hence, the only black hole solution of the
four-dimensional Einstein-Maxwell theory is the Kerr-Newman black hole.
This result precludes the possibility that a black hole possesses higher
multipole moments (\eg a mass quadrupole or a charge dipole) that are
not completely fixed by the values of the conserved charges. Another
consequence is that the microscopic states that are responsible for the
large degeneracy implied by the Bekenstein-Hawking entropy, are
invisible at the level of the classical gravitational theory.

We will review here the recent discovery that five-dimensional
black holes exhibit qualitatively new properties not shared by their
four-dimensional siblings. In short,
non-spherical horizon topologies are possible, and conventional notions
of black hole uniqueness do not apply \cite{ER}. This, and other recent
developments ---notably, the study of the interplay between black holes
and black strings in theories with compact dimensions \cite{revbhbs}---
have made plainly clear that the physics of higher-dimensional black
holes is largely {\it terra incognita}, and have prompted its
exploration in earnest.

Although the higher-dimensional version of the Schwarzschild solution
was found long ago \cite{tangher}, it was not until 1986, with the
impetus provided by the development of string theory, that the
higher-dimensional version of the Kerr solution was constructed by Myers
and Perry (MP) \cite{MP}. Given that the Kerr black hole solution is
unique in four dimensions, it may have seemed natural to expect black
hole uniqueness also in higher dimensions.

We know that, at least in five dimensions, and very likely in $D\geq 5$
dimensions, this is not the case. A heuristic argument that suggests the
possibility of black holes of non-spherical topology is the following.
Take a neutral black string in five dimensions, constructed as the
direct product of the Schwarzschild solution and a line, so the geometry
of the horizon is ${\bf R}\times S^2$. Imagine bending this string to
form a circle, so the topology is now $S^1\times S^2$. In principle this
circular string tends to contract, decreasing the radius of the $S^1$,
due to its tension and gravitational self-attraction. However, we can
make the string rotate along the $S^1$ and balance these forces against
the centrifugal repulsion. Then we end up with a neutral rotating {\it
black ring}: a black hole with an event horizon of topology $S^1 \times
S^2$. Ref.~\cite{ER} obtained an explicit solution of
five-dimensional vacuum general relativity describing such an object.
This was not only an example of non-spherical horizon topology, 
but it also turned out to be a counterexample to black hole uniqueness.

The discovery that black hole uniqueness is violated in higher
dimensions was greeted with surprise but, with the benefit of hindsight,
what we should really regard as surprising is that black hole uniqueness
{\it is} valid in four dimensions. The no-hair property of
four-dimensional black holes was regarded as very surprising at the time
of its discovery, and the realization that it does not extend to higher
dimensions serves to emphasize this.

\medskip
\noindent {\em The Forging of the Ring}

Originally, an investigation of {\it static} solutions \cite{weyl}
combined with educated guesswork (which gave the solutions in
\cite{dgkt,CE,composite} from which black rings are obtained via
analytic continuation) was the way to the rotating black ring solutions
of \cite{ER,RE}. Recently, these solutions have been rederived in a
systematic manner via solution-generating techniques
\cite{japanrings,sto}. The same techniques have also given vacuum black
rings with rotation on the $S^2$ but without rotation along the $S^1$
ring circle \cite{phir} ---however, an educated guess has independently given the
same solution in a much more manageable form \cite{PF}. In contrast, the
charged supersymmetric and non-supersymmetric black rings have, from the
start, been constructed in a more systematic way.

\medskip
\noindent {\em A note on nomenclature}

A black ring is defined to be a $D$-dimensional black hole for which the
topology of (a spatial cross-section of) the event horizon is $S^1
\times S^{D-3}$. (So far, such solutions are only known in $D=5$ but we
allow for the possibility that such objects may exist for $D > 5$.)
Often one wishes to distinguish black rings from topologically spherical
black holes, which are often abbreviated to simply ``black holes". Black
rings are also black holes, but the context should eliminate any
possible confusion.

\medskip
\noindent {\em Outline of this review}

A main difficulty in understanding the black ring solutions appears to
be the somewhat unfamiliar coordinate system in which they take their
simplest known form. Therefore, we devote the next section
\ref{sec:ringcoords} to a derivation and explanation of these
coordinates. This may also be useful for obtaining adapted coordinates
in other settings. Section \ref{sec:vacsol} studies the neutral black
ring and how it gives rise to non-uniqueness in five dimensions. Section
\ref{sec:chardip} introduces black rings as charged sources of gauge
fields. The particularly interesting case of supersymmetric black rings
is analyzed in section \ref{sec:susyrings}. Section \ref{sec:micro}
examines the ideas underlying the microscopic description of black rings
in string and M theory. In section \ref{sec:otherstuff} we discuss other
developments related to the role of black rings in string/M theory. The
final section briefly addresses some open issues and ideas for future
work.

\section{Ring coordinates}
\label{sec:ringcoords}

The rotation group in four spatial dimensions, $SO(4)$, contains two
mutually commuting $U(1)$ subgroups, meaning that it is possible to have
rotation in two independent rotation planes. In order to see this point
more clearly, consider four-dimensional flat space and group the four
spatial coordinates in two pairs, choosing polar coordinates for each of
the two planes,
\beq\label{twoplanes}
x^1=r_1 \cos\phi,\quad x^2=r_1\sin\phi,\quad x^3=r_2\cos\psi,\quad
x^4=r_2\sin\psi\,.
\eeq
Rotations along $\psi$ and $\phi$ generate two independent angular
momenta $J_\psi$ and $J_\phi$. We will describe rings extending along
the $(x^3,x^4)$-plane, and rotating along $\psi$, thus giving rise to
non-vanishing $J_\psi$. 

\begin{figure}[!t]
\begin{picture}(0,0)(0,0)
{\small
\put(260,128){$x=-1$}
\put(145,255){$y=-1$}
\put(210,241){$y=\mathrm{const}$}
\put(308,204){$x=\mathrm{const}$}
}
\end{picture}
\centering{\psfig{file=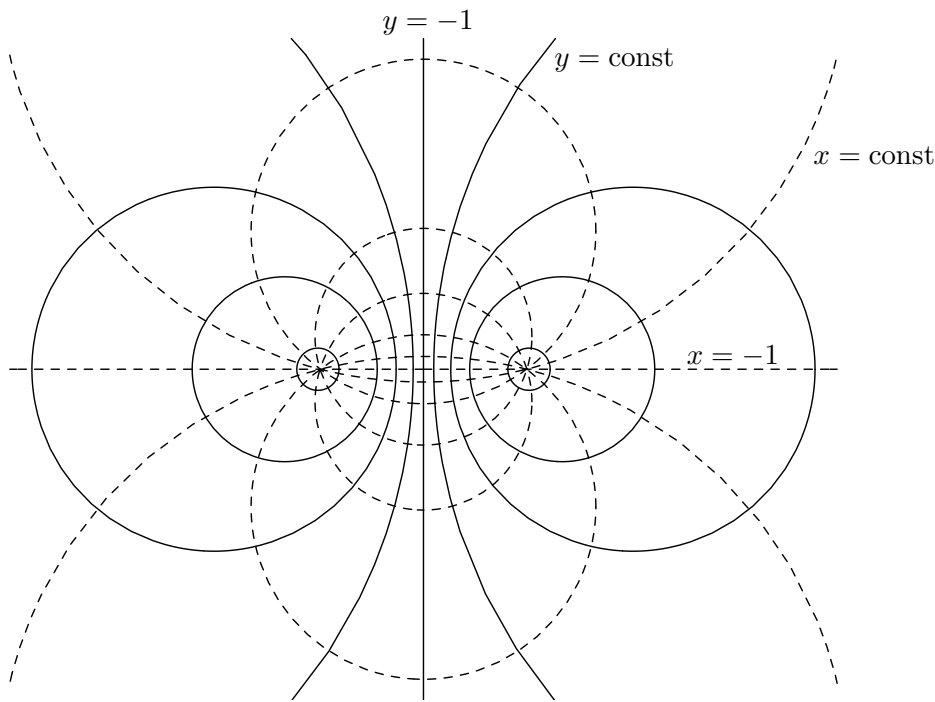,width=11cm}} 
\vspace{4ex}
\caption{\small 
Ring coordinates for flat four-dimensional space, on a section at
constant $\phi$ and $\psi$ (and $\phi+\pi$, $\psi+\pi$). Dashed circles
correspond to spheres at constant $|x|\in [0,1]$, solid circles to
spheres at constant $y\in [-\infty,-1]$.
Spheres at constant $y$ collapse to zero size at the location of
the ring of radius $R$, $y=-\infty$. The disk bounded by the ring is an
axis for $\partial_\phi$ at
$x=+1$.
}\label{fig:Ringcoords}
\end{figure}

As is often the case in General Relativity, it is very convenient to
work in adapted coordinates. A general idea to find them is to begin by
constructing a foliation of flat space in terms of the equipotential
surfaces of the field created by a source resembling the black hole one
is seeking\footnote{In \cite{HO1} a similar approach is followed to
obtain coordinates suitable for black holes and black strings on a
Kaluza-Klein circle.}. It turns out that, instead of considering the
equipotential surfaces of a {\it scalar} field sourced by a ring, it is
more convenient to work with the equipotential surfaces of a {\it 2-form
potential} $B_{\mu\nu}$. Thus we regard the ring as a circular string
that acts as an electric source of the 3-form field strength $H=dB$,
which satisfies the field equation
\beq\label{twoformeq}
\partial_\mu(\sqrt{-g}H^{\mu\nu\rho})=0\,
\eeq
outside the ring source. Let us write four-dimensional flat space in the
coordinates of \reef{twoplanes}
\beq\label{flatr1r2}
d{\bf x}_{\it 4}^2=dr_1^2+r_1^2 d\phi^2+dr_2^2+r_2^2 d\psi^2 \,.
\eeq
It is easy to construct the solution of \reef{twoformeq} for a circular
electric source at $r_1=0$, $r_2=R$ and $0\leq \psi<2\pi$ using methods
familiar in classical electrodynamics, as (see \cite{EMT})
\beqa\label{bfield}
B_{t\psi}&=&\frac{R}{2\pi}\int_0^{2\pi} d\psi
\frac{r_2\cos\psi}{r_1^2+r_2^2 +R^2 -2Rr_2\cos\psi}\nonumber\\
&=& -\frac{1}{2}\left(1-\frac{R^2+r_1^2+r_2^2}{\Sigma} \right)
\eeqa
where 
\beq
\Sigma=\sqrt{(r_1^2+r_2^2+R^2)^2-4R^2r_2^2}\,.
\eeq 
We can as easily find the electric-magnetic (Hodge) dual of this field.
In five spacetime dimensions, $\ast H= F=dA$ where $A$ is a one-form
potential, so the dual of an electric string is a magnetic monopole---in
this case a circular distribution of monopoles. Note that surfaces of
constant $A_\phi$ will be orthogonal to surfaces of constant $B_{t\psi}$. For the
dual of the field \reef{bfield} one finds
\beq\label{afield}
A_\phi=-\frac{1}{2}\left(1+\frac{R^2-r_1^2-r_2^2}{\Sigma} \right)\,.
\eeq
Now define coordinates $y$ and $x$ that correspond to constant values of  
$B_{t\psi}$ and $A_\phi$, respectively. A convenient choice is
\beq\label{yxr12}
y=-\frac{R^2+r_1^2+r_2^2}{\Sigma}\,,\qquad x=\frac{R^2-r_1^2-r_2^2}{\Sigma}\,,
\eeq
with inverse
\beq\label{r12yx}
r_1=R\frac{\sqrt{1-x^2}}{x-y}\,,\qquad r_2=R\frac{\sqrt{y^2-1}}{x-y}\,.
\eeq
Observe that the coordinate ranges are 
\beq\label{xyrange}
-\infty\leq y\leq -1\,, \qquad -1\leq x\leq 1
\eeq
with $y=-\infty$ corresponding to the location of the ring source, and
asymptotic infinity recovered as $x\to y\to-1$. The axis of rotation
around the $\psi$ direction, $r_2=0$ (actually not a line but a plane) is
at $y=-1$, and the axis of rotation around $\phi$, $r_1=0$, is divided
into two pieces: $x=1$ is the disk $r_2\leq R$, and $x=-1$ is its
complement outside the ring, $r_2\geq R$. In these coordinates the flat
metric \reef{flatr1r2} becomes 
\beq\label{flatxy}
d{\bf x}_{\it 4}^2=\frac{R^2}{(x-y)^2}\left[
(y^2-1)d\psi^2+\frac{dy^2}{y^2-1}
+\frac{dx^2}{1-x^2}
+(1-x^2)d\phi^2\right]\,.
\eeq
This is depicted in fig.~\ref{fig:Ringcoords}, where we present a
section at constant $\psi$ and $\phi$ (as well as the antipodal
sections at $\psi+\pi$, $\phi+\pi$ for greater clarity). 

We can rewrite this same foliation of space in a manner that is
particularly appropriate
in the region near the ring. Define
coordinates $r$ and $\theta$ as
\beq\label{rtheta}
r=-\frac{R}{y}\,,\qquad \cos\theta=x\,,
\eeq
with
\beq
0\leq r\leq R\,,\qquad 0\leq \theta \leq \pi \,.
\eeq
The flat metric \reef{flatxy} becomes
\beq\label{flatrtheta}
d{\bf x}_{\it 4}^2=\frac{1}{\left(1+\frac{r\cos\theta}{R}\right)^2}
\left[
\left(1-\frac{r^2}{R^2}\right)R^2 d\psi^2+\frac{dr^2}{1-r^2/R^2}
+ r^2\left(d\theta^2+\sin^2\theta d\phi^2\right)
\right]\,.
\eeq
Note that the apparent singularity\footnote{Which, not by accident,
may remind some readers of the deSitter horizon.} at $r=R$ actually
corresponds to the $\psi$-axis of rotation.

It is now manifest that the surfaces at constant $r$, \ie
constant $y$, have ring-like topology $S^2\times S^1$,
where the $S^2$ is parametrized by $(\theta,\phi)$ and the $S^1$ by
$\psi$. The black rings will have their horizons (and ergosurfaces) at
constant values of $y$, or $r$, and so their topology will be clear in these
coordinates.

In the flat space \reef{flatrtheta}, the $S^2$'s at constant $r$ are
actually metrically round spheres, centered at $r_2=R/\sqrt{1-r^2/R^2}$,
with radius $r/\sqrt{1-r^2/R^2}$. Due to the overall prefactor in
\reef{flatrtheta}, this is not quite obvious. However,
for small spheres $r\ll R$ the prefactor is
$\approx 1$, and $r$ and $\theta$ recover their conventional
interpretation as the radius and polar angle on the spheres\footnote{
The surfaces at constant $|x|$ and
constant $\phi$ are also round spheres, centered at
$r_1=R|x|/\sqrt{1-x^2}$ and with radius $R/\sqrt{1-x^2}$.
}.
Hence the coordinates $r$ and $\theta$ are natural in the region
of small $r$, but they look bizarre at larger distances. In particular 
asymptotic infinity corresponds to $r\cos\theta=-R$. The coordinates
$(x,y)$ are physically opaque, but they preserve a symmetry under exchange
$x\leftrightarrow y$ that is otherwise obscured, and allow for more
compact expressions.

Incidentally, $\Sigma^{-1}$ solves the Laplace equation for a ring
sourcing a scalar field, $\nabla^2 \Sigma^{-1}=0$. Since
$\Sigma^{-1}=(x-y)/2R^2$, we see that surfaces of constant scalar
potential do not correspond to constant $x$ nor $y$, except in the limit
of large negative $y$ ($r\ll R$) where $\Sigma^{-1}\simeq -y/2R^2=1/2Rr$. It is
possible to construct coordinates adapted to surfaces of constant
$\Sigma$ and their gradient surfaces, but the form of the black ring
solutions becomes somewhat more complicated. The $(x,y)$ coordinates,
being adapted to the two-form potential $B$, also facilitate greatly the
analysis of solutions with gauge dipoles, in particular of
supersymmetric black rings.

\section{Neutral Black Ring}
\label{sec:vacsol}

\subsection{Spacetime geometry}

The metric for the black ring geometry preserves most of the basic
structure of \reef{flatxy}, but now it contains additional functions
that encode the non-zero curvature produced by the black ring. In
$(x,y)$ coordinates these functions admit a particularly simple form, as
they can be written as linear functions of $x$ and $y$.

The solution has been given in three related forms in \cite{ER},
\cite{HT,EE}, and \cite{RE}, the latter two forms being more convenient than the
original one. The form given in \cite{RE} appears to be more
fundamental, since black rings with a dipole, or with rotation in the
$S^2$ \cite{PF}, are more naturally connected to this version of the solution. The
metric is\footnote{If we denote quantities in \cite{EE} with a hat, then
the relationship is
$x =\frac{\hat x-\hat\lambda}{1-\hat\lambda\hat x}$, 
$y =\frac{\hat y-\hat\lambda}{1-\hat\lambda\hat y}$, 
$(\phi,\psi)=\frac{1-\hat\lambda\hat\nu}{\sqrt{1-\hat\lambda^2}}(\hat\phi,\hat\psi)$,
$\nu=\frac{\hat\lambda-\hat\nu}{1-\hat\lambda\hat \nu}$,
 and
$\lambda=\hat\lambda$.}
\beqa\label{neutral}
ds^2&=&-\frac{F(y)}{F(x)}\left(dt-C\: R\:\frac{1+y}{F(y)}\:
d\psi\right)^2\nonumber\\[2mm]
&&+\frac{R^2}{(x-y)^2}\:F(x)\left[
-\frac{G(y)}{F(y)}d\psi^2-\frac{dy^2}{G(y)}
+\frac{dx^2}{G(x)}+\frac{G(x)}{F(x)}d\phi^2\right]\,,
\eeqa
where
\beq\label{fandg}
F(\xi)=1+\lambda\xi,\qquad G(\xi)=(1-\xi^2)(1+\nu\xi)\,,
\eeq
and
\beq\label{coeff}
C=\sqrt{\lambda(\lambda-
\nu)\frac{1+\lambda}{1-\lambda}}\,.
\eeq
The dimensionless parameters $\lambda$ and $\nu$ must lie in the range 
\beq\label{lanurange}
0< \nu\leq\lambda<1\,.
\eeq 
When both $\lambda$ and $\nu$ vanish we recover flat spacetime in the
form \reef{flatxy}. $R$ sets the scale for the solution, and $\lambda$ and
$\nu$ are parameters that characterize the shape and rotation velocity
of the ring, as we shall clarify presently. 


Although we shall work primarily with the metric in these coordinates
and parameters, since they allow for more compact expressions, it is
instructive to also consider the $(r,\theta)$ coordinates introduced in
\reef{rtheta}, as well as to redefine the parameters $(\nu,\lambda)\to
(r_0,\sigma)$ as
\beq\label{ringpar}
\nu =\frac{r_0}{R}\,,\qquad \lambda=\frac{r_0\cosh^2\sigma}{R}\,.
\eeq
The solution becomes perhaps uglier,
\beqa\label{ringrtheta}
ds^2&=&-\frac{\hat f}{\hat g}\left(dt-
r_0\sinh\sigma\cosh\sigma\sqrt{\frac{R+r_0\cosh^2\sigma}{R- r_0\cosh^2\sigma}}\:
\frac{\frac{r}{R}-1}{r\hat f}\:R\:d\psi\right)^2 \nonumber\\
&&
+
\frac{\hat g}{\left(1+\frac{r\cos\theta}{R}\right)^2}
\left[
\frac{f}{\hat f}\left(1-\frac{r^2}{R^2}\right)\, R^2d\psi^2
+
\frac{dr^2}{(1-\frac{r^2}{R^2})f}
+ 
\frac{r^2}{g}\,d\theta^2
+
\frac{g}{\hat g}\, r^2 \sin^2\theta\, d\phi^2
\right]
\eeqa
where
\beq\label{ffs}
f=1-\frac{r_0}{r}\,,\qquad \hat f=1-\frac{r_0\cosh^2\sigma}{r}\,,
\eeq
and 
\beq\label{ggs}
g=1+\frac{r_0}{R} \cos\theta\,,\qquad
\hat g=1+\frac{r_0\cosh^2\sigma}{R} \cos\theta \,.
\eeq
Consider the limit 
\beq\label{thinring}
r,\,r_0,\,r_0\cosh^2\sigma\ll R
\eeq
in which $g,\,\hat g \approx 1$, and redefine $\psi=z/R$. Then
\reef{ringrtheta} becomes exactly the metric for a boosted black string,
extended along the direction $z$, and with boost parameter $\sigma$. The
horizon is at $r=r_0$, and absence of conical singularities requires
that $\psi$ be identified with period $2\pi$ so the string is
periodically identified with
radius $R$: $z\sim z+2\pi R$. Hence the limit \reef{thinring} corresponds
to taking the ring radius $R$ much larger than the ring thickness $r_0$,
and focusing on the region near the ring $r\sim r_0$. 

This gives precise
meaning to the heuristic construction of a black ring as a boosted black
string bent into circular shape. It also allows to give an
approximate interpretation to $\lambda$ and $\nu$. According to
\reef{ringpar}, the parameter $\nu$ measures the ratio between the
radius of the $S^2$ at the horizon, $r_0$, and the radius of the ring
$R$. So smaller values of $\nu$ correspond to thinner rings. Also,
$\lambda/\nu$ is a measure of the speed of rotation of the ring. More
precisely, $\sqrt{1-(\nu/\lambda)}$ can be approximately identified with
the local boost velocity $v=\tanh\sigma$. 

We now turn to a general analysis of the metric in the form \reef{neutral}.
The coordinates $x$ and $y$ vary within the same range as in
\reef{xyrange} and are interpreted in essentially the same manner as we
saw in the previous section. An important difference, though, is that
for general values of the parameters in \reef{lanurange} the orbits of
$\partial/\partial\psi$ and $\partial/\partial\phi$ do not close off
smoothly at their respective axes, but in general have conical
singularities there. To avoid them at $x=-1$ and $y=-1$ the angular
variables must be identified with periodicity
\beq\label{period0}
\Delta\psi=\Delta\phi=4\pi\frac{\sqrt{F(-1)}}{|G'(-1)|}=
2\pi\frac{\sqrt{1-\lambda}}{1-\nu}\,.
\eeq
To avoid also a conical singularity at $x=+1$ we must have $\Delta\phi=
2\pi\sqrt{1+\lambda}/(1+\nu)$. This is compatible with \reef{period0}
only if we take the two parameters $\lambda$, $\nu$, to satisfy
\beq\label{equil0}
\lambda=\frac{2\nu}{1+\nu^2}\,.
\eeq
Fixing $\lambda$ to this value leaves only two independent parameters in
the solution, $R$ and $\nu$. In fact this is as expected on physical
grounds: given, say, the mass and the radius of the ring, the angular
momentum must be tuned so that the centrifugal force balances the
tension and self-attraction of the ring, thus leaving only two free parameters.
Demanding the absence of conical singularities, as in \reef{equil0},
actually corresponds to the condition that the system is balanced
without any external forces.

Note that in terms of the boost parameter $\sigma$ introduced in
\reef{ringpar}, and in the limit of thin rings \reef{thinring}, the
equilibrium value from \reef{equil0} becomes
\beq
|\sinh\sigma|\to 1\,
\eeq
or equivalently, the velocity $|v|\to 1/\sqrt{2}$. This happens to be the value
of the boost that makes the ADM pressure of the black string vanish,
$T_{zz}=0$ \cite{EE}. The latter holds for all known black rings
in equilibrium, including dipole rings \cite{RE}. Presumably it applies
in more generality, \eg for higher-dimensional black rings and black
objects with more complicated horizon topology.

With the choices \reef{period0} and \reef{equil0} for the parameters the
solution becomes
asymptotically flat as $x\to y\to -1$. Since the geometry is distorted
by the presence of curvature, in order to go to manifestly
asymptotically flat coordinates we have to modify \reef{r12yx} slightly.
We set
\beq\label{asympinf}
\tilde r_1=\tilde R\frac{\sqrt{2(1+x)}}{x-y}\,,\quad\tilde r_2=\tilde
R\frac{\sqrt{-2(1+y)}}{x-y}\,,\quad {\tilde
R}^2=R^2\frac{1-\lambda}{1-\nu}\,,\quad
(\tilde\psi,\tilde\phi)=\frac{2\pi}{\Delta\psi}(\psi,\phi)
\eeq
(note that we are taking $x$ and $y$ close to $-1$, and that
$\tilde\psi,\tilde\phi$ have canonical periodicity). Then \reef{neutral}
asymptotes to the flat space metric \reef{flatr1r2}, now with `tilded'
coordinates $\tilde r_{1,2}, \tilde\psi,\tilde\phi$.

Note that $F(y)$ vanishes at $y=-1/\lambda$. Nevertheless, it is easy to
check that the metric and its inverse are smooth there. This locus
corresponds to a timelike surface in spacetime at which
$\partial/\partial t$ changes from timelike to spacelike, \ie it is an
{\it ergosurface}. A spatial cross section of this surface has topology
$S^1 \times S^2$.

At $y=-1/\nu$ the metric becomes singular, but we can show that this is
only a coordinate singularity by the transformation $(t,\psi)\to
(v,\psi')$ as
\beq
dt=dv-C R\frac{1+y}{G(y)\sqrt{-F(y)}}dy\,,\qquad
d\psi=d\psi'+\frac{\sqrt{-F(y)}}{G(y)}dy\,.
\eeq
In these coordinates the metric is 
\beqa
  ds^2 &=& -\frac{F(y)}{F(x)}\left( dv-C R \frac{1+y}{F(y)} d\psi' \right)^2\nonumber \\
&& + \frac{R^2}{(x-y)^2} F(x)
\left[
-\frac{G(y)}{F(y)} {d\psi'}^2
+2\frac{d\psi'\,dy}{\sqrt{-F(y)}}
+\frac{dx^2}{G(x)}
+\frac{G(x)}{F(x)} d\phi^2
\right] \,,
\eeqa
which is manifestly regular at $y=-1/\nu$. Let
\be
V = \frac{\partial}{\partial t} +
\Omega\frac{\partial}{\partial\tilde\psi} = \frac{\partial}{\partial v}
+ \Omega \frac{\partial}{\partial \tilde \psi'},
\ee
where $\tilde\psi' = (2\pi/\Delta \psi)\psi'$ and
\beq\label{omega}
\Omega= \frac{1}{R}\sqrt{\frac{\lambda-
\nu}{\lambda(1+\lambda)}}\,.\\[2mm]
\eeq
Then $V$ is null at $y=-1/\nu$ and $V_\mu dx^\mu$ is a positive multiple
of $dy$, from which it follows that $y=-1/\nu$ is a Killing horizon with 
angular velocity $\Omega$. In the limit \reef{thinring} of a thin ring
we recover $\Omega R=\tanh\sigma$.

This horizon has spatial topology $S^1\times S^2$, although the $S^2$ is
distorted away from perfect sphericity. At $y=-\infty$ the invariant
$R_{\mu\nu\sigma\rho}R^{\mu\nu\sigma\rho}$ blows up, which corresponds
to an inner spacelike singularity . 

The Myers-Perry black
hole with rotation in a single plane is contained within the family of
solutions \reef{neutral} as the particular limit\footnote{The limit is
much less singular in the coordinates of \cite{HT,EE}.} in which $R\to
0$, and $\lambda,\nu\to 1$, while
maintaining fixed the parameters $a$, $m$,
\beq\label{tompbh}
m=\frac{2R^2}{1-\nu}\,,\qquad a^2=2R^2\frac{\lambda-\nu}{(1-\nu)^2}\,,
\eeq
changing coordinates
$(x,y)\to (r,\theta)$,
\beqa
x&=&-1+2\left(1-\frac{a^2}{m}\right)\frac{R^2\cos^2\theta}{r^2-(m-
a^2)\cos^2\theta}\,,\nonumber\\
y&=&-1-2\left(1-\frac{a^2}{m}\right)\frac{R^2\sin^2\theta}{r^2-(m-
a^2)\cos^2\theta}\,,
\eeqa
and rescaling
$
(\psi,\phi)\to \sqrt{\frac{m-a^2}{2R^2}}\;(\psi,\phi)
$
so they now have canonical periodicity $2\pi$. Then we recover the metric
\beq\label{MPbh}
ds^2=-\left(1-\frac{m}{\Sigma}\right)\left(dt+\frac{m
a\sin^2\theta}{\Sigma-
m}\:d\psi\right)^2+\Sigma\left(\frac{dr^2}{\Delta}+d\theta^2\right)
+\frac{\Delta\sin^2\theta}{1-
m/\Sigma}\:d\psi^2+r^2\cos^2\theta\:d\phi^2\,,
\eeq
\beq
\Delta\equiv r^2-m+a^2\,,\qquad \Sigma\equiv
r^2+a^2\cos^2\theta
\eeq
of the MP black hole rotating in the $\psi$ direction. The extremal
limit $m=a^2$ of the MP black hole actually corresponds to the same
nakedly singular solution obtained as $\nu\to 1$ in \reef{neutral}.

\subsection{Physical magnitudes and Non-Uniqueness}

To demonstrate the absence of uniqueness for this family of solutions we
need their two conserved charges: the mass and spin. These are obtained
by examining the metric near asymptotic infinity, $x\to y\to-1$, in the
more conventional coordinates of \reef{asympinf}, and comparing to the
linearized gravity analysis in \cite{MP}.\footnote{
Although note that the solutions of \cite{MP} are all rotating in a {\it
negative} sense. Positive rotation corresponds to $g_{t\psi}$ negative
near infinity.}
We find
\beq\label{mass}
M=\frac{3\pi R^2}{4G }\frac{\lambda}{1-
\nu}\,,
\eeq
\beq\label{spin}
J=\frac{\pi R^3}{2G }\frac{\sqrt{\lambda(\lambda-
\nu)(1+\lambda)}}{(1-\nu)^2}\,.
\eeq
The horizon area and temperature (from the surface gravity $\kappa=2\pi T$) are
\beq\label{area}
{{\cal A}_H}=8\pi^2 R^3
\frac{\nu^{3/2}\sqrt{\lambda(1-
\lambda^2)}}{(1-\nu)^2(1+\nu)}\,,
\eeq
\beq\label{tempN}
T=\frac{1}{4\pi
R}(1+\nu)\sqrt{\frac{1-
\lambda}{\lambda\nu(1+\lambda)}}\,.
\eeq

To analyze the physical properties of the solutions it is convenient to
first fix the overall scale. Instead of fixing $R$, which has no
invariant meaning, we shall fix the mass $M$. The solutions can then be
characterized by dimensionless magnitudes obtained by dividing out an
appropriate power of $M$ or of $G M$ (which has dimension (length)$^2$).
So we define a dimensionless ``reduced spin" variable $j$, conveniently
normalized as
\beq\label{etadef}
j^2\equiv \frac{27\pi}{32G}\frac{J^2}{M^3}\,,
\eeq
($j^2$ is often a more
convenient variable than $j$), as well as a reduced area of the horizon,
\beq\label{zetadef}
{a_H}\equiv\frac{3}{16}\sqrt{\frac{3}{\pi}}\frac{{\cal A}_H}{(G M)^{3/2}}\,.
\eeq
Above we argued that a black ring
at equilibrium, \ie satisfying \reef{equil0}, has
only one independent dimensionless parameter.
Therefore at equilibrium the reduced area and
spin, ${a_H}$ and
$j$, must be related. Using the results above, this
can be expressed in parametric form as
\beq\label{zetaeta}
{a_H}= 2\sqrt{\nu(1-\nu)}\,,\qquad
j^2= \frac{(1+\nu)^3}{8\nu}\qquad \mathrm{(black\: ring)}\,,
\eeq
with $0<\nu\leq 1$. 

For the spherical MP black hole \reef{MPbh} the corresponding relation
can be found in \cite{MP}, or also by taking the limit from the general
ring solution as explained above in \reef{tompbh}. The result is
\beq\labell{zetaetabh}
{a_H}=2\sqrt{2(1-j^2)}\qquad \mathrm{(MP\: black\: hole)}\,.
\eeq

\begin{figure}[!th]
\begin{picture}(0,0)(0,0)
{\small 
\put(78,120){MP black hole}
\put(160,50){thin black ring}
\put(60,20){fat black ring}
}
{\large
\put(160,-25){$j^2$}
\put(-25,100){$a_H$}
}
\small{
\put(-2,-4){0}
\put(106,-12){$\frac{27}{32}$}
\put(130,-10){1}
\put(-18,176){$2\sqrt{2}$}
\put(-5,60){$1$}
}
\end{picture}
\centering{\psfig{file=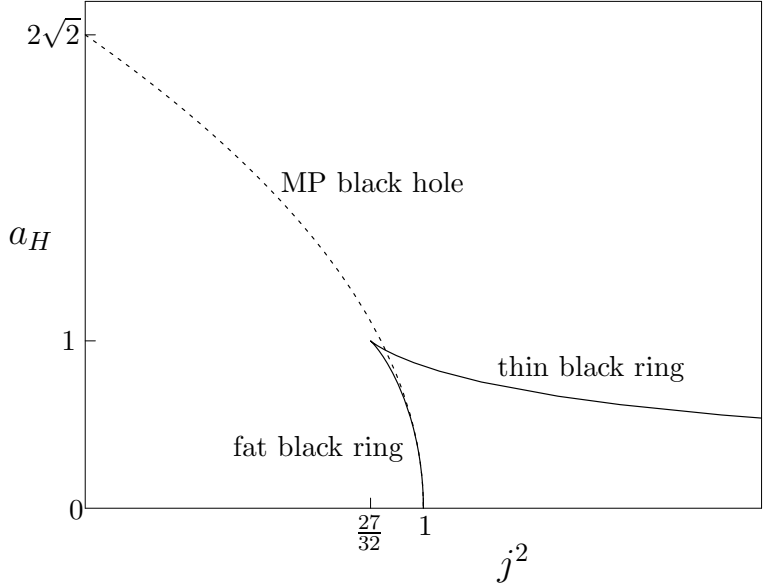,width=9cm}} 
\vspace{4ex}
\caption{\small 
Horizon area $a_H$ vs\ (spin)$^2$  $j^2$, for given mass, for the neutral
rotating black ring (solid) and MP black hole (dotted). There are two
branches of black rings, which branch off from the cusp at 
$(j^2,a_H) = (27/32,1)$, and which are dubbed ``thin" and ``fat"
according to their shape. When $27/32 < j^2<1$ we find the {\it Holey Trinity:} 
three different solutions ---two black rings and one MP black hole--- with the same 
dimensionless parameter $j$ (\ie with the
same mass and spin). The minimally spinning ring, with $j^2=27/32$,
has a regular non-degenerate horizon, so it is not an extremal solution.
Other interesting features are:
At $j^2={a_H}^2=8/9$ 
the curves intersect and we find a MP black hole and a (thin) black ring
both with the same
mass, spin and area. The limiting solution at $(j^2,
a_H)=(1,0)$ is a naked singularity. Rapidly spinning black rings, 
$j^2\to\infty$, become thinner and their area decreases as ${a_H}\sim
1/(j\sqrt{2})$.
}\label{fig:ajneu}
\end{figure}

The curves \reef{zetaeta} and \reef{zetaetabh} are plotted and described
in figure \ref{fig:ajneu}. The plot exhibits several unusual features.
For instance, contrary to what happens for rotating black holes in four
dimensions, and for the MP black hole in five dimensions, the angular
momentum of the black ring (for fixed mass) is bounded below, but not
above. 
But the most striking feature is that in the range $27/32\leq j^2<1$
there exist one MP black hole and two black rings all with the same
values of the mass and the spin. Since the latter are the only conserved
quantities carried by these objects, we have an explicit violation of
black hole uniqueness.

It is sometimes asserted that the existence of the black ring implies
{\em per se} a violation of black hole uniqueness. However, we do not
know of any a priori argument why the respective ranges of $j$ for MP black
holes and black rings should overlap\footnote{At least not an argument
within classical relativity. But perhaps one might argue from
thermodynamics that the phases in the diagram fig.~\ref{fig:ajneu}
should exhibit the generic `swallowtail' structure, as indeed they do.}.
Only through examination of the explicit solutions do we see that they
are, respectively, $j^2<1$ and $j^2\geq 27/32$, and so indeed they do
overlap---but then only rather narrowly so! Observe also that it is not
possible to recover a notion of uniqueness by fixing the horizon
topology, since there can be two black rings with the same $M$ and $J$.

\section{Charges and Dipoles}
\label{sec:chardip}

For topologically spherical black holes (\eg the Kerr-Newman solution), the
combination of electric charges and rotation gives rise to associated
magnetic dipoles, which do not violate uniqueness since they do not
provide parameters independent of the conserved charges. 

Black rings can carry conserved gauge charges (the first example was
obtained in \cite{HE}). More remarkably, they can also support gauge
dipoles that are {\it independent} of all conserved charges, in fact
they can be present even in the absence of
any gauge charge. So, generically, these dipoles entail continuous
violations of uniqueness.  This is a much more drastic effect than the
discrete, three-fold non-uniqueness that we have found for neutral
rings.

The charges and dipoles of black rings actually provide
the basis to interpret them as objects in string/M-theory. The
five-dimensional supergravities of which these black rings are solutions
are then most conveniently viewed as dimensional reductions of 
eleven-dimensional supergravity, the low-energy limit of M-theory, as we review next.

\subsection{Dimensional Reduction to Five Dimensions}
\label{subsec:redto5}

We start with eleven-dimensional supergravity, whose bosonic fields are
the metric and a 3-form potential ${\cal A}$ with 4-form field strength
${\cal F} = d{\cal A}$. The action is
\be
I_{11} = \frac{1}{16 \pi G_{11}} \int \left( R_{11} \star_{11} 1 -
\frac{1}{2} {\cal F} \wedge \star_{11}{\cal F} - \frac{1}{6} {\cal F}
\wedge {\cal F} \wedge {\cal A}
\right) \,,
\ee
where $R_{11}$ and $\star_{11}$ denote the eleven-dimensional Ricci
scalar and Hodge dual, respectively. We shall be interested in a
five-dimensional supergravity theory obtained by dimensional reduction
on $T^6$ using the Ansatz\footnote{More generally, one can compactify on
a Calabi-Yau
three-fold, leading to $N=1$ supergravity in five-dimensions coupled to
a certain number of vector multiplets (and hypermultiplets, which we
need not consider). The results below are easily
generalized to this case.}
\beqa
ds_{\it 11}^2 &=&  ds^2_{\it 5} + X^1 \left( dz_1^2 +
dz_2^2 \right) + X^2 \left( dz_3^2 + dz_4^2 \right) +
X^3 \left( dz_5^2 + dz_6^2 \right),\nonumber \\
\mbox{${\cal A}$} &=& A^1 \wedge dz_1 \wedge dz_2 + A^2 \wedge dz_3 \wedge dz_4 +
A^3 \wedge dz_5 \wedge dz_6 \,.
\label{11sol}
\eeqa
It is assumed that nothing depends on the coordinates $z^i$
parametrizing the $T^6$, so we can regard $ds^2_{\it 5}$, $X^i$ and
$A^i$ as a five-dimensional metric, scalars, and vectors respectively.
We assume that the scalars $X^i$ obey a constraint
\beq
\label{eqn:constr}
 X^1 X^2 X^3 = 1.
\eeq
This ensures that the $T^6$ has constant volume, which guarantees that
the metric $ds^2_{\it 5}$ is the five-dimensional Einstein-frame metric.
The eleven-dimensional action reduces to the action of five-dimensional
$U(1)^3$ supergravity:
\beq
 I_5 = \frac{1}{16 \pi G_5} \int \left( R \star 1 - G_{ij} dX^i \wedge
\star dX^j - G_{ij} F^i \wedge \star F^j - \frac{1}{6} C_{ijk} F^i
\wedge F^j \wedge A^k \right) \,,
\label{5action}
\eeq
where $G_{ij} = \frac{1}{2} {\rm diag} \left( (X^1)^{-2},(X^2)^{-2},(X^3)^{-2} \right)$,
$C_{ijk}=1$ if $(ijk)$ is a permutation of $(123)$ and $C_{ijk}=0$
otherwise, and the Maxwell field strengths are $F^i = dA^i$. In five
dimensions, we can define conserved electric charges for asymptotically
flat solutions by
\beq
\mathbf{Q}_i = \frac{1}{16\pi G_5} \int_{S^3} (X^i)^{-2} \star_5 F^i,
\label{charge}
\eeq
where the integral is evaluated at spatial infinity. Using standard
techniques (\eg \cite{gibbonshull}), it can be shown that any
appropriately regular solution of this theory satisfies the BPS
inequality
\be
\label{eqn:bpsineq}
 M \ge |\mathbf{Q}_1| + |\mathbf{Q}_2| + |\mathbf{Q}_3|\,,
\ee
where $M$ is the ADM mass. A solution is said to be supersymmetric if it
saturates this inequality. Examining the eleven-dimensional field
strength makes it clear that the electric charge $\mathbf{Q}_i$ that
couples to $F^i$ arises from M2-branes wrapped on the internal $T^6$,
\eg $F^1$ is sourced by M2-branes wrapping the $12$ cycle of $T^6$
etc. The charges are quantized in terms of the wrapping numbers of the
M2-branes as 
\be
N_i=\left(\frac{4G_5}{\pi}\right)^{1/3} \mathbf{Q}_i\,.
\label{quantN}
\ee
We can also map these solutions to a U-duality frame that is convenient for
microscopic analysis, namely, as solutions for a D1-D5-brane intersection
with momentum running along their common direction. To
this effect, dimensionally reduce the eleven-dimensional solution above
on (say) the $z^6$ direction to give a solution of ten-dimensional type
IIA supergravity. Performing T-dualities in the $z^5,z^4,z^3$ directions
then gives a solution of type IIB supergravity with metric
\beq
 ds^2 = - (X^3)^{1/2} ds_{\it 5}^2
          + (X^3)^{-3/2} \left(dz + A^3\right)^2
      + X^1 (X^3)^{1/2} d{\bf z}_4^2,
\eeq
where the $T^5$ is parametrized by the coordinates $z\equiv z^5$, ${\bf
z}_4\equiv(z^1, z^2, z^3, z^4)$. Note
that the circle parametrized by $z$ may be non-trivially fibered over
the five-dimensional spacetime. We shall refer to this direction as the
Kaluza-Klein circle.
The other non-zero IIB fields are
\beq
  e^{2\Phi} = \frac{X^1}{X^2} \,, \qquad
F_{(3)} = \left( X^1 \right)^{-2} \star_5 F^1 + F^2 \wedge \left(
dz+A^3 \right)\,,
\label{2bfields}\eeq
where $\Phi$ is the dilaton, $F_{(3)}$ the Ramond-Ramond 3-form field
strength, and $\star_5$ denotes the Hodge dual with respect to the
five-dimensional metric.
These formulae allow any
solution of five-dimensional $U(1)^3$ supergravity to be uplifted to a
solution of type IIB supergravity. Examining the RR 3-form reveals that
the electric charges that couple to $F^1$ and $F^2$ arise from D5-branes
wrapped on $T^5$ and D1-branes wrapped around the $z$-circle
respectively. The appearance of $A^3$ in the metric reveals that $F^3$
is electrically sourced by momentum ($P$) around the KK $z$-circle.

\subsection{Dipoles}
\label{sec:dipole}

\begin{figure}[th]
\begin{picture}(0,0)(0,0)
{
\put(140,18){$S^2$}
}
\end{picture}
\centering{\psfig{file=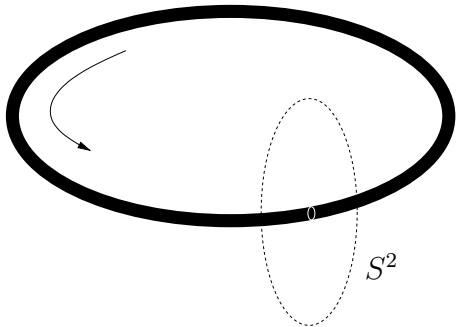,width=6cm}} 
\caption{\small The dipole $q_i$ measured from the magnetic
flux of $F^i$ across an $S^2$
that encloses a section of the string. An azimuthal angle has been
suppressed in the picture, so
the $S^2$ is represented as a circle.}
\label{fig:ringdipole}
\end{figure}

The possible presence of dipoles on a black ring is most easily
understood by recalling our discussion of the field of a circular string
in Section \ref{sec:ringcoords}. There we saw that a circular string
gives rise to an electric field $B_{t\psi}$, whose magnetic dual
$A_\phi$ is sourced by a circular distribution of magnetic monopoles.
So, in addition to the charges $\mathbf{Q}_i$ defined in \reef{charge},
the topology of a black ring allows to define `dipole charges' $q_i$ as
we would do for a magnetic charge,
\be
q_i  =
\frac{1}{2 \pi} \int_{S^2} F^i\,,
\label{dipole}
\ee
by performing the integral on a surface $S^2$ that links the ring
once, see figure \ref{fig:ringdipole}. The field \reef{afield}
corresponds to a unit dipole.
 
However, even if there is a local distribution of charge, the total
magnetic charge is zero: in order to compute the magnetic charge in five
dimensions, one has to specify a two-sphere that encloses a point of the
ring, {\it and} a vector tangent to the string. So there are opposite
magnetic charges at diametrically opposite points of the ring, which
justifies the analogy to a dipole. The magnetic charges
can be completely annihilated by contracting the size of the ring to
zero, so $q_i$ is not a conserved quantity. Equivalently, in the dual
electric picture, the field $B_{t\psi}$ is not associated to any
conserved charge since $\psi$ parametrizes a contractible cycle.

One might also attempt to define the dipole
from the asymptotic fall off of the gauge fields, which, from
\reef{afield}, is 
\beq
A_\phi \to  q \frac{R^2 r_1^2}{(r_1^2+r_2^2)^2}+O(r_1^2+r_2^2)^{-2}\,.
\eeq
However, if there are electric charges present, the magnetic dipole
field will have a component induced by the rotation, which does not
contribute to non-uniqueness. It is therefore preferable to use the
definition \reef{dipole} to characterize the dipole
intrinsic to the source.

Dipole rings are therefore specified by the independent physical
parameters $(M,J,q_i)$. The $q_i$ are non-conserved, classically
continuous parameters. So they imply continuous violations of
non-uniqueness in five dimensions. Rotating black rings with these
dipoles were conjectured to exist in refs.~\cite{tubular,harv,EE}, and
were first found in \cite{RE}.

In the embedding into M-theory described in the previous section, the
black ring is made of M5-branes, with four worldvolume directions
wrapping 4-cycles of the internal $T^6$ and the fifth direction being
the circular ring direction. The charge $q_i$ is then quantized as
\be
n_i=
\left(\frac{\pi}{4G_5}\right)^{1/3} q_i\,,
\label{quantn}\ee
corresponding to the wrapping number of the M5-branes around the ring
circle.

The solution for the dipole black rings of \reef{5action} contains, in
addition to the parameters that the neutral solution already has, three
new ones, $\mu_i$. The three dipoles $q_i$ are then functions of the
$\mu_i$, and when all $\mu_i=0$ we recover the neutral solution
\reef{neutral}. The explicit form of the metric is
\beqa\label{magnetic}
ds_5^2&=&-\frac{F(y)}{F(x)}\frac{H(x)}{H(y)}
\left(dt-C \: R\:\frac{1+y}{F(y)}\: d\psi\right)^2\\[3mm]
&&+\frac{R^2}{(x-y)^2}\: F(x)H(x)H(y)^2\left[
-\frac{G(y)}{F(y)H(y)^3}d\psi^2-\frac{dy^2}{G(y)}
+\frac{dx^2}{G(x)}+\frac{G(x)}{F(x)H(x)^3}d\phi^2\right]\,.\nonumber
\eeqa 
The functions $F$ and $G$ are as in \reef{fandg}, $C$ is defined by \reef{coeff}, and 
\beq\label{hfactor}
H(\xi)= \left[H_1(\xi)H_2(\xi)H_3(\xi)\right]^{1/3}\,,
\eeq
with \beq
H_i(\xi)=1-\mu_i\xi\,. 
\eeq 
The gauge potentials are
\beq\label{Ai}
A^i=C_i \:R\:\frac{1+x}{H_i(x)}\:d\phi\,,
\eeq
and the scalars
\beq
X^i=\frac{H(x) H_i(y)}{H(y) H_i(x)}\,. 
\eeq
$C_i$ is as in \reef{coeff} but with $\lambda\to -\mu_i$.
Since the field \reef{Ai} is purely magnetic, it makes no contribution
to the Chern-Simons term in the action.

The parameters $\lambda$ and $\nu$ vary in the same ranges as in the neutral case
\reef{lanurange}, while
\beq\label{murange}
0\leq \mu_i<1\,.
\eeq
Expressions for the mass etc of this solution can be found in \cite{RE}.
The addition of dipoles to a ring has several effects on its
dynamics. The dipole increases the self-attraction between opposite
points along the ring. For given black ring mass, if the dipole is
non-vanishing the angular momentum is not only bounded below but also
bounded above. Also, for fixed mass and angular momentum, the addition
of a dipole reduces the area and the temperature of the black ring. When
the three dipoles are present, the dipole ring has an outer and an inner
horizon, and an upper bound on the magnitude of $q$ is obtained when the
two horizons coincide ($\nu=0$) and the ring becomes extremal. This ring
has a non-singular horizon of finite area and vanishing temperature.
However, it is {\it not} supersymmetric, since in the absence of any
conserved charges $\mathbf{Q}_i$ the BPS bound \reef{eqn:bpsineq} cannot
be saturated.

More surprisingly, dipoles feature in the first
law
\beq\label{1stlaw}
dM=\frac{Td\mathcal{A}_H}{4G_5}+\Omega dJ+\Phi^i d\mathbf{Q}_i+ \phi^i
dq_i\,.
\eeq
Here $\phi^i$ is defined (up to convenient normalization) as the
difference in the dipole potential at infinity and at the
horizon. When $\mathbf{Q}_i=0$ this equation was obtained in \cite{RE}
from the explicit form of the dipole ring solutions. However, as noted
in \cite{CH}, the last term appears to be at odds with
conventional derivations of the first law, which seem to allow only
conserved charges into it. The resolution of this puzzle lies in the
impossibility to define the dipole potential, using a single patch, such
that it is simultaneously regular at the rotation axis $y=-1$ and at the
horizon. As a result, a new surface term enters the first law, giving
precisely \reef{1stlaw} \cite{CH} (see also \cite{ast}).

\section{Supersymmetric Black Rings}
\label{sec:susyrings}

\subsection{The solution and its properties}

The possible existence of supersymmetric black rings was suggested in
\cite{BK,bena} based on thought experiments involving supersymmetric
black holes and supertubes. The subsequent discovery of supersymmetric
black ring solutions grew out of parallel studies of charged black rings
\cite{EE,EEF} and of a program to classify supersymmetric solutions of
five-dimensional $N=1$ supergravity. It turns out that there is a
canonical form for such solutions \cite{gghpr,gr2}, with the necessary
and sufficient conditions for supersymmetry reducing to simple-looking
equations on a four-dimensional ``base space". The first supersymmetric
black ring solution \cite{EEMR1} was obtained by solving these equations
for minimal 5D supergravity, taking the base space to be flat space
written in ring coordinates as in (\ref{flatxy}). This was subsequently
generalized to $U(1)^3$ supergravity by three independent groups
\cite{BW,EEMR2,JGJG2}.\footnote{In fact, the same method yields black
ring solutions of $U(1)^n$ supergravity \cite{EEMR2,JGJG2}.} The
solution is: 
\beqa
\label{eqn:5dsol}
ds_{\it 5}^2 &=& -(H_1 H_2 H_3)^{-2/3} (dt + \omega)^2 +
(H_1 H_2 H_3)^{1/3} d{\bf x}_{\it 4}^2,\nonumber \\
A^i &=&  H_i^{-1}(dt +\omega) +
\frac{q_i}{2} \left[(1+y) d\psi + (1+x) d\phi \right], \label{5sol} \\
X^i &=& H_i^{-1} (H_1 H_2 H_3)^{1/3} \,, \nonumber
\eeqa
where $d{\bf x}_{\it 4}^2$ is the flat base space that we encountered in
\reef{flatxy},
the functions $H_i$ are
\beqa
H_1 &=& 1 + \frac{Q_1 - q_2 q_3 }{2R^2} (x-y) -
\frac{q_2 q_3}{4 R^2} (x^2 - y^2),\nonumber \\
H_2 &=& 1 + \frac{Q_2 - q_3 q_1 }{2R^2} (x-y) -
\frac{q_3 q_1}{4 R^2} (x^2 - y^2),\\
H_3 &=& 1 + \frac{Q_3 - q_1 q_2 }{2R^2} (x-y) -
\frac{q_1 q_2}{4 R^2} (x^2 - y^2),\nonumber
\label{eqn:Hi}
\eeqa
and $\omega = \omega_\phi d\phi + \omega_\psi d\psi$ with
\beqa
 \omega_\phi &=&  \frac{1}{8 R^2} (1-x^2) \left[ q_1 Q_1 + q_2 Q_2 + q_3 Q_3 -
   q_1 q_2 q_3 \left( 3 + x + y
   \right) \right], \\
 \omega_\psi &=& -\frac{1}{2} (q_1 + q_2 + q_3) (1+y)  + \frac{1}{8R^2}
(y^2-1) \left[
q_1 Q_1 + q_2 Q_2 + q_3 Q_3 - q_1 q_2 q_3 \left( 3 + x + y \right)
\right].\nonumber
\label{eqn:omegas}\eeqa
The coordinate ranges are as they would be for the metric
(\ref{flatxy}). 

The solution depends on the seven parameters $q_i$, $Q_i$ and $R$. The
$q_i$ are dipole charges defined by (\ref{dipole}) with the integral
taken over a surface of constant $t$, $y$ and $\psi$. $Q_i$ are
proportional to the conserved charges:
\be
\mathbf{Q}_i  = \frac{\pi}{4G_5} Q_i,
\ee
and $R$ is a length scale corresponding to the radius of the ring with
respect to the base space metric. In the limit $R \rightarrow \infty$
with the charge densities $Q_i/R$ fixed, the solution reduces to a black
string solution obtained in \cite{bena} (the change of coordinates and limit
required are essentially the same as in the neutral case \reef{ringrtheta}). 
Supersymmetry implies that the mass is fixed by saturation of the BPS
inequality (\ref{eqn:bpsineq}). The three Killing fields generate a ${\bf R}
\times U(1) \times U(1)$ isometry group, just as for nonsupersymmetric
black rings.

The most obviously novel feature of this solution is the fact that
$\omega_\phi \ne 0$: the solution rotates in both the $\phi$ and $\psi$
directions. The angular momenta are
\beqa
J_{\phi} &=& \frac{\pi}{8G_5} \left(q_1 Q_1 + q_2
Q_2 + q_3 Q_3 - q_1 q_2 q_3 \right),\\
\qquad J_{\psi} &=&
\frac{\pi}{8G_5} \left[2R^2 (q_1 +q_2 + q_3) + q_1 Q_1 + q_2 Q_2 + q_3
Q_3 - q_1 q_2 q_3 \right].\nonumber
\eeqa
Note that the parameter $R$ is determined by $J_{\psi} - J_{\phi}$ and
the dipoles.

Many supersymmetric solutions suffer from causal pathologies such as
closed causal curves (CCCs) \cite{gghpr}. The necessary and sufficient
condition for the above solution to be free of closed causal curves for
$y \ge -\infty$ is \cite{EEMR2}
\be
  2 q^2 L^2 \equiv 2 \sum_{i < j} \mathcal{Q}_i q_i \mathcal{Q}_j q_j - \sum_i
  \mathcal{Q}_i^2 q_i^2 - 4 R^2 q^3 \sum_i q_i \ge 0,
\label{eqn:noctcs}
\ee
where we have defined
\be\label{qcalQ}
q = (q_1 q_2 q_3)^{1/3}, \qquad \mathcal{Q}_1 = Q_1 - q_2 q_3, \qquad
\mathcal{Q}_2 = Q_2 - q_3 q_1,
\qquad \mathcal{Q}_3 = Q_3 -
 q_1 q_2.
\ee
If the inequality in (\ref{eqn:noctcs}) is strict then the solution has
an event horizon at $y \rightarrow -\infty$ \cite{EEMR2,JGJG2}. Just as
for non-supersymmetric rings, $\psi$ and $t$ are not good coordinates on
the horizon and have to be replaced by new coordinates $\psi'$ and $v$.
For supersymmetric rings, the $\phi$ rotation implies that $\phi$ is
also not a good coordinate, however $\chi \equiv \phi-\psi$ is. The
geometry of a spacelike section of the horizon is
\be
ds_{H}^2 = L^2 d{\psi'}^2 + \frac{q^2}{4}
\left(d\bar{\theta}^2+\sin^2\bar{\theta} d\chi^2\right),
\ee
where $x=\cos \bar\theta$ as before. So the horizon is geometrically a
product of a circle of radius $L$ and a two-sphere of radius $q/2$. The
entropy can be calculated from the horizon area:
\be\label{susyS}
 S = \frac{ \pi^2 \, L \,q^2}{2G_5} = 2\pi\sqrt{\frac{c \hat{q}_0}{6}},
\ee
where, using the quantized charges $N_i$ and $n_i$ in \reef{quantN},
\reef{quantn},
\be
 c = 6 n_1 n_2 n_3,
\ee
and
\be\label{hatq0}
\hat{q}_0 = \frac{n_1 n_2 n_3}{4} + \frac{1}{2} \left( \frac{N_1
N_2}{n_3} + \frac{N_2 N_3}{n_1} + \frac{N_1N_3}{n_2} \right) -
\frac{1}{4 n_1 n_2 n_3} \left[(N_1 n_1)^2 + (N_2 n_2)^2 + (N_3 n_3)^2
\right]-J_{\psi}.
\ee
The reason for writing the entropy is this rather odd form will become
apparent when we discuss the microscopic interpretation of
supersymmetric black rings.

Finally, we note that the angular velocities of the horizon of a
supersymmetric black ring vanish, as is necessarily the case for a
supersymmetric, asymptotically flat black hole \cite{gmt}.

\subsection{Non-uniqueness}

Before the discovery of supersymmetric black rings, the only know
supersymmetric black hole solution of five-dimensional supergravity was
the so-called BMPV black hole \cite{bmpv}. This solution can be obtained
as a limit of the supersymmetric black ring solution. To this end,
consider the change of coordinates
\be
  \rho \cos{\Theta} = r_1, \qquad \rho \sin{\Theta} = r_2\,, 
  \label{rhoTheta}
\ee
where $r_1,r_2$ were defined in (\ref{r12yx}) and $0\leq \rho<\infty$,
$0\leq \Theta\leq \pi/2$. In these coordinates the flat base space
metric is
\beqa
  d{\bf x}_4^2 = d\rho^2
  + \rho^2 (d\Theta^2 + \sin^2{\Theta}d\psi^2+\cos^2{\Theta}d\phi^2)
\, .
\label{rhoThetabase}
\eeqa
The form of the above solution in these coordinates is given in
\cite{EEMR2}. To obtain the BMPV solution we take the limit $R
\rightarrow 0$ with the coordinates and other parameters held fixed. The
solution becomes
\be
 H_i = 1 + \frac{Q_i}{\rho^2}, \qquad A^i = H_i^{-1} (dt + \omega),
\ee
\be
  \omega_\phi = \frac{4G_5J}{\pi} \frac{\cos^2{\Theta}}{\rho^2}
  \, ,~~~~
  \omega_\psi = \frac{4G_5J}{\pi} \frac{\sin^2{\Theta}}{\rho^2} \,,
\ee
where
\be
  J=\frac{\pi}{8G_5}
  \left[ q_1 Q_1 + q_2 Q_2 + q_3 Q_3 - q_1 q_2 q_3 \right] \, .
\ee
This solution is determined by four parameters: $Q_i$ and $J$. The
former retain their interpretation as conserved M2-brane charges. The
latter determines the angular momenta: $J_\phi=J_\psi=J$. The solution
has an event horizon at $\rho=0$ of topology $S^3$. The solution is much
more symmetrical than the supersymmetric black ring solution: it has
isometry group ${\bf R} \times U(1) \times SU(2)$ \cite{gmt}, although this is
not manifest in the above coordinates.

Since the angular momenta of the BMPV black hole are equal and those of
a supersymmetric black ring are always unequal, it follows that one can
always distinguish these two types of supersymmetric black hole by
comparing their conserved charges. Nevertheless, the conserved charges
of a supersymmetric black ring can still be made arbitrarily close to
those of a BMPV black hole by taking $R$ small enough. 

Although supersymmetric black rings cannot carry the same conserved
charges as a BMPV black hole, the fact that dipole charges are required
to specify them entails a violation of black hole uniqueness in exactly
the same way as for non-supersymmetric dipole rings. Seven parameters
are required to specify the solution but there are only five independent
conserved charges, namely $\mathbf{Q}_i$, $J_\phi$ and $J_\psi$. Hence
there is a continuous violation of black hole uniqueness even for
supersymmetric black holes. However, if one takes charge quantization
into account then this violation of uniqueness is rendered finite
\cite{EEMR2}.

A more extreme violation of black hole uniqueness was proposed in
\cite{BW}. It was realized in \cite{BW} that 
the problem of finding supersymmetric solutions can be reduced to
specifying appropriate sources for certain harmonic functions on the
base space. Physically, these sources describe M2-branes with both
worldvolume directions wrapped on the internal torus, and M5-branes with
four worldvolume directions wrapped on the torus, \ie ``M2-particles"
and ``M5-strings" in the five noncompact directions. Choosing the
sources to correspond to a circular loop of M5-strings with a constant
density of M2-particles distributed around the string leads directly to
the supersymmetric black ring solution above \cite{BW}. One can also
construct more general solutions for which the loop of M5 branes is not
(geometrically) a circle or the density of M2 branes not constant
\cite{BW,BWW}. However, it turns out that such non-uniform solutions do
not admit smooth horizons so they do not describe black holes \cite{HR}.
Nevertheless, one can still assign an entropy to them formally and it
may be possible to make sense of them in string theory. 

\subsection{Multi-ring solutions}
\label{subsec:multi}

In $D=4$, there exist solutions of Einstein-Maxwell theory describing a
static superposition of several extremal Reissner-Nordstrom black holes
\cite{hartle}. Physically, these solutions exist because there is a
cancellation of gravitational attraction and electrostatic repulsion
between the holes. This is often the case for supersymmetric systems (of
which these solutions are an example \cite{gibbonshull}). It is
therefore natural to ask whether there exist solutions describing
multiple supersymmetric black rings or superpositions of supersymmetric
black rings with BMPV black holes. It turns out that such solutions do
indeed exist \cite{JGJG1,JGJG2}.

The simplest way to understand these solutions is to imagine
constructing them using the method of \cite{BW} and prescribing sources
corresponding to multiple rings. However, in order to obtain the
corresponding solutions in a form sufficiently explicit to analyze in
detail, a different approach is required. Following \cite{JGJG1} we can
introduce new coordinates $\chi = \phi - \psi$, $\alpha = -\phi-\psi$,
$\bar{\theta} = \Theta/2$ and $\bar{r} = \rho^2/(4\ell)$, where $\Theta$
and $\rho$ are defined in (\ref{rhoTheta}) and $\ell$ is an arbitrary
length. The flat metric (\ref{rhoThetabase}) becomes
\be
d{\bf x}_4^2  = H^{-1} \left(d\alpha + \cos \bar{\theta} d\chi \right)^2 + H
\left(d\bar{r}^2 + \bar{r}^2 d\bar{\theta}^2 + \bar{r}^2\sin^2
\bar{\theta} d\chi^2 \right),
\ee
where $H = \ell/\bar{r}$. This flat metric is a special case of a more
general family of Ricci-flat metrics known as {\it Gibbons-Hawking}
metrics \cite{gibbonshawking}:
\be
ds^2 = H^{-1} \left( d\alpha + {\bf w} \right)^2 + H \delta_{ij} dx^i
dx^j,
\ee 
with $\partial/\partial \alpha$ a Killing field, ${\bf w} \equiv w_i
dx^i$ obeys $\nabla \times {\bf w} = \nabla H$ (where $\nabla_i =
\partial_i$), which implies that $H$ is harmonic on ${\bf R}^3$:
$\nabla^2 H = 0$. It was shown in \cite{gghpr,JGJG2} that the general
supersymmetric solution with a Gibbons-Hawking base space for which
$\partial/\partial \alpha$ extends to a space-time symmetry is specified
by harmonic functions on ${\bf R}^3$ (of which $H$ is one). The
supersymmetric black ring solution discussed above is such a solution.
It turns out that, with the exception of $H$, the harmonic functions for
black ring solutions all have a single pole at a certain point on the
negative $z$-axis in ${\bf R}^3$ with the position related to the
``radius" $R$ of the ring.

The multi-ring solutions of \cite{JGJG1,JGJG2} are obtained by taking
the harmonic functions to have more general sources\footnote{Although
still with $H=\ell/\bar{r}$, \ie a flat base space.} with poles at
several points in ${\bf R}^3$, each corresponding to a different ring. If
these poles all lie along the $z$-axis then the resulting solution
preserves the same $U(1) \times U(1)$ symmetry on the base space as a
single black ring. This implies that, on the base space, each ring
corresponds to a circle centred on the origin and lying either in the 12
plane or the 34 plane in the coordinates of (\ref{twoplanes}). (Whether
it is the 12 plane or the 34 plane is dictated by whether the relevant
pole in the harmonic functions is on the negative or positive $z$-axis).
The solutions admit regular event horizons and appear free of causal
pathologies provided certain restrictions on the parameters are
satisfied.\footnote{ A possible explanation for the these restrictions
is the following. Supersymmetry guarantees cancellation of gravitational
and electromagnetic forces, but not
of forces arising from spin-spin interactions. The extra condition might
arise from requiring that these interactions vanish.}

More general solutions, in which the poles do not all lie on the
$z$-axis can also be constructed \cite{JGJG1,JGJG2}. These preserve only
a single $U(1)$ symmetry on the base space and correspond to rings
centered on the origin but no longer restricted to the 12 or 34 planes.
They appear to be free of pathologies close to individual rings but it
is not known what extra conditions are required for these solutions to
be well-behaved globally. 

Note that if one takes a multi-ring system and shrinks one of the rings
down to zero radius then one obtains a solution describing a BMPV black
hole sitting at the common centre of the remaining rings
\cite{JGJG1,JGJG2}. For the case of a single ring and a single BMPV
black hole, such a solution can be written down using the original ring
coordinates of (\ref{flatxy}) \cite{BW}. 

Finally, we note that the above construction can be generalized by
replacing the flat base space with a more general Gibbons-Hawking space.
A particularly interesting choice is (self-dual, Euclidean) Taub-NUT
space, which corresponds to $H = 1 + \ell/\bar{r}$. This has the same
topology as ${\bf R}^4$ but differs geometrically. Surfaces of constant
$\bar{r}$ have $S^3$ topology but, viewing $S^3$ as a $S^1$ bundle over
$S^2$, the radius of the $S^1$ approaches a constant as $\bar{r}
\rightarrow \infty$ whereas the radius of the $S^2$ grows as $\bar{r}$.
Hence solutions with Taub-NUT base space obey Kaluza-Klein, rather than
asymptotically flat, boundary conditions. The method of
\cite{gghpr,JGJG2} can be used to obtained solutions describing BMPV
black holes \cite{gsy} and (multiple concentric) supersymmetric black
rings \cite{EEMR3,gsy2,bkw} in Taub-NUT space. After Kaluza-Klein
reduction, the latter solutions correspond to the 4D multi-black hole
bound states obtained in \cite{denef}. This ``4D-5D connection" allows
one to extend recently discovered relations between 4D black holes and
topological string theory to 5D black holes \cite{gsy}.

\section{Microscopics of Black Rings}
\label{sec:micro}

The microscopic description of black holes in string theory is typically
based on the dynamics of a configuration of branes that has the same set
of charges as the black hole. Black rings can carry both conserved and
dipole charges: depending on which of the two sets one puts the stress
on, the description is rather different. But they are related: the
conserved-charge-based description is the ultraviolet (UV)
completion of the dipole-based one, which describes only the physics of
the system at the lowest energies, \ie the infrared (IR). 

Both theories are two-dimensional sigma-models, and in the extremal
limit where the momentum is chiral the
entropy follows from the Cardy
formula
\beq\label{cardy}
S=2\pi\sqrt{\frac{c \hat q_0}{6}}\,.
\eeq
The central charge, $c$, and the momentum available to
distribute among chiral oscillators, $\hat q_0$, differ in each
description (even if we refer to the same object). The two-dimensional
sigma-model can be regarded as an ``effective string" and the
descriptions differ in what we take this string to be:
\begin{itemize}

\item The IR theory has the effective string extending along the $S^1$
direction of the ring, so we view the ring as a circular string. This
was first proposed and applied to extremal non-supersymmetric black
rings in \cite{RE}. Its application to supersymmetric black rings gives
an impressive match of the statistical and Bekenstein-Hawking entropies
\cite{BK2,CGMS}.

\item In the UV theory the effective string direction extends along a
sixth-dimension orthogonal to the ring. Thus we view the ring as a
tube---more properly, a supertube, or an excitation of it. This was
first proposed for two-charge black rings in \cite{EE}, and developed
for supersymmetric three-charge black rings in \cite{BK2}.

\end{itemize}

As we will see, each description has its virtues and shortcomings.

\subsection{IR theory: Black rings as circular strings}

This is based on the worldvolume theory of the branes that carry the
dipoles, which have one worldvolume direction along the ring circle. The
microscopic theory for the straight string limit of the ring is then
applied to a circular ring of finite radius. So far, none of the
proposed IR theories can distinguish between a ring and a KK
compactified string, \ie they work to the extent that finite radius
corrections to the entropy cancel out.

For a black
ring with a finite horizon in the extremal limit, a convenient
description is obtained in terms of a triple intersection of M5-branes,
with momentum running along the ring, as we have seen in sections
\ref{sec:chardip} and \ref{sec:susyrings}. If the six compact directions
of
space are small, then the low energy dynamics is described by a
$(0,4)$-supersymmetric $1+1$ sigma-model at the intersection of the
branes. Following \cite{MSW}, the central charge $c$ of the theory is
given by the number of moduli that parametrize the deformations of the
(smoothed) intersection of branes, and is proportional to the number of
branes of each kind. A detailed calculation gives
\beq
c_{IR}=6 n_1 n_2 n_3\,.
\eeq
The supergravity and sigma-model descriptions are valid when the volume
of the six-dimensional internal space (in 11D Planck units) is,
respectively, $V_6\ll c_{IR}$, or $V_6\gg c_{IR}$. Additionally, one
requires the radius of the ring $R\gg V_6^{1/6}$ (to
reduce to a sigma-model), and $V_6\gg 1$ (to neglect quantum
corrections to supergravity) \cite{MSW}.

Conserved charges corresponding to M2 branes are obtained by turning on
worldvolume fluxes. For the moment, we set these fluxes to zero. Then
the ring carries only M5 dipoles and angular momentum $J_\psi$, which in
the absence of fluxes is identified with the effective string momentum
$\hat q_0$. At finite radius, this is not a supersymmetric solution even
if the momentum is chiral. This corresponds to a black ring solution that is
the extremal limit of the dipole ring of \reef{magnetic}. It is
straightforward to check that the resulting microscopic degeneracy
formula
\beq 
S=2\pi\sqrt{n_1 n_2 n_3 J_\psi }\,
\eeq 
correctly describes the entropy of the black ring in the straight string
limit $R\to\infty$, which is no more than the known entropy match for
the black string \cite{MSW}. Ref.~\cite{RE} showed that the model does
even better, since it also captures correctly the leading corrections of
the black ring entropy in a $1/R$ expansion.\footnote{Bear in mind that $R$ is
not an independent parameter but is fixed by the other charges so in a
$1/R$ expansion we take some combination of charges to be large.} These
corrections already include the effects of the self-interaction among
different points along the ring ---an effect of the finite ring radius.
These effects become too strong in subleading corrections, however, and
it is not clear how to account for them.

Supersymmetric black rings provide a better behaved system: finite
radius effects appear to be absent from the entropy at any radius,
although this has not been explained microscopically and so the
conclusions are less rigorous than would be desired. To saturate the BPS
bound, supersymmetric black rings necessarily carry conserved M2
charges, with integer brane numbers $N_i$; in the microscopic picture,
we turn on fluxes on the worldvolume of the M5 branes. These fluxes also
give rise to momentum zero-modes that contribute to the total momentum
$q_0$, so this is no longer equal to the momentum available to
non-zero-mode oscillators $\hat q_0$. Instead the relationship is
\cite{MSW,CGMS}
\beq
\hat{q}_0 = q_0 +
\frac{1}{2} \left( \frac{N_1 N_2}{n_3} + \frac{N_2 N_3}{n_1} +
\frac{N_1 N_3}{n_2} \right) - \frac{1}{4 n_1n_2n_3} \Big(
(N_1  n_1)^2 + (N_2 n_2)^2 + (N_3 n_3)^2 \Big) + \frac{n_1n_2n_3}{4} \,.
\label{qhat}
\eeq
Comparing to \reef{hatq0} we see that the choice $q_0=-J_\psi$ yields a
perfect match of the microscopic entropy \reef{cardy} to the
Bekenstein-Hawking entropy \reef{susyS}. 
The last term in \reef{qhat} is a zero-point correction to the
momentum and as we see is necessary to find perfect agreement with the
entropy of the black ring. 

In the analysis of \cite{BK2} the identification of
parameters is slightly different. It can be checked that $\hat{q}_0$
in \reef{hatq0} can alternatively be written as
\beq
\hat{q}_0 = -J_\psi+J_\phi +
\frac{1}{2} \left( \frac{{\cal N}_1 {\cal N}_2}{n_3} + \frac{{\cal N}_2
{\cal N}_3}{n_1} +
\frac{{\cal N}_1 {\cal N}_3}{n_2} \right) - \frac{1}{4 n_1n_2n_3} \Big(
({\cal N}_1  n_1)^2 + ({\cal N}_2 n_2)^2 + ({\cal N}_3 n_3)^2 \Big) \,,
\label{qhat2}
\eeq
where ${\cal N}_1=N_1-n_2 n_3$, and the obvious permutations for ${\cal
N}_{2,3}$. Ref.~\cite{BK2} proposes that ${\cal Q}_i$, 
instead of
$Q_i$, is the quantity to be identified with the M2-brane charge at the
source, so the actual microscopic M2-brane numbers are ${\cal N}_i$
instead of $N_i$. One must note, though, that there is no known invariant
definition of charge that justifies this choice \cite{HR}. If one then
equates $q_0=-J_\psi+J_\phi$, and ignores the zero-point term in
\reef{qhat}, the entropy is reproduced. This match uses crucially the
fact that the entropy is given by the quartic invariant of the U-duality
group $E_7$ of M-theory on CY$_3 \times S^1$ at low energies, \ie the
theory that describes a KK compactified black string, so finite radius
effects are again ignored. A possible rationale why the entropy of the
compactified KK string and the black ring should agree is given in
\cite{bkw} via consideration of black rings in Taub-NUT. By varying the
modulus corresponding to the KK radius, one can interpolate between the
compactified black string and the black ring. Since the entropy is
moduli-invariant it should remain the same for both limits. 

It is striking, and not at all well understood, that these two different
calculations reproduce exactly the entropy of the black ring.
Although the match between the entropies is remarkable, both of these microscopic
descriptions clearly leave many points obscure by being
unable to say anything about finite radius effects. Besides the problems
already mentioned, the calculation in \cite{CGMS} does not explain why
the M2 charges $Q_i$ are bounded below by the dipoles so that ${\cal
Q}_i>0$. Moreover, the microscopic picture would seem to place no
restrictions on the angular momentum $J_\psi$ nor the ring radius $R$.
However, these cannot be varied independently if the ring is to remain
in equilibrium. The role of the second angular momentum ($J_\phi$ in
\cite{CGMS}, $J_\psi+J_\phi$ in \cite{BK2}), which
does not appear anywhere in the entropy formulas, is also unclear (some
ideas are discussed in \cite{EEMR2,EM}). Note that this
second angular momentum is fixed by the other charges, and it should be
possible to calculate it in the microscopic theory. The result should
agree with supergravity because angular momentum is quantized and hence
not renormalized. However, the proposals of \cite{BK2,CGMS} would appear
to ascribe a vanishing value to this angular momentum.

But perhaps a more important deficiency, inherent to the description, is
that it does not allow to say anything about the microscopic
significance of black hole non-uniqueness. In a sense, the dipole-based
IR theory looks too closely at the ring, and by focusing on the
string-like aspects of the ring, it cannot describe the spherical black
hole. To be able to view both black objects from a unified perspective,
we have to step back and observe them from a greater distance (so, by
AdS/CFT duality, we go to the ultraviolet), where the conserved charges
play the dominant role.

\subsection{UV theory: Black rings as supertubes }

The dynamics is now determined by the worldvolume theory on the branes
that carry the conserved charges. In principle this CFT can describe
both spherical black holes and black rings with the same charges as
different phases of the theory, with the dipoles acting as order
parameters. Depending on the phase, the theory has a different flow to
the IR. While the spherical black hole phase is exactly conformal, the
black ring induces a non-trivial flow to the theory of the previous
subsection. One can check that the central charge of the IR theory is
indeed less than that of the UV theory \cite{BK2,KL}, which is
quantitative evidence in favour of this picture. The
supersymmetries in one of the chiral sectors of the (4,4) UV theory are
broken along the flow, so the IR theory has (0,4) supersymmetry.

It is convenient to pass to a different U-duality frame.
The five-dimensional supersymmetric black hole is best understood by first
uplifting it to a black string in six dimensions, and viewing it as an
intersection of D1 and D5-branes that carry momentum along this sixth,
common direction, as discussed near the end of
subsec.~\ref{subsec:redto5}. Again, at low energies the dynamics is
captured by a $1+1$ CFT with central charge
\beq
c_{UV}=6N_1N_2
\eeq
where $N_1$, $N_2$ are the numbers of D1 and D5 branes. 

This sigma-model CFT is well-understood only at a point in moduli space
where its target space is a symmetric orbifold of $N_1 N_2$ copies of
the internal four-manifold. Roughly speaking, at this point the theory
is `free'. The supergravity description, instead, corresponds to a
deformation of the theory away from it (a `strong coupling' regime).
However, the symmetric orbifold theory seems to capture correctly many
of the features of black holes with D1-D5 charges.

At the orbifold point the CFT contains twisted sectors. Pictorially, the
maximally twisted sector corresponds to a long effective string, of
length $N_1 N_2$ times the length of the physical circle that the string
wraps (which we take to be equal to one). The energy gap of momentum
excitations is then smallest $\sim 1/N_1N_2$. The untwisted sector can
be regarded as containing a number $N_1 N_2$ of short effective strings,
each of unit length. Momentum excitations have large gap $\sim O(1)$.
There are also partially twisted sectors.

To describe the spherical black hole with D1-D5-P charges, we put the
string in the maximally twisted sector. The linear momentum $P$, which
in this case is one of the three conserved charges, is carried by both
bosonic and fermionic excitations along the effective string. The
fermionic excitations can also carry a polarization in the transverse
directions (as R-charge in the CFT). If there are $q_J$ such oscillators
polarized in the same direction, they give rise to a self-dual angular
momentum
\beq
J_\psi=J_\phi=J=\sqrt{\frac{c_{UV} q_J}{6}}\,.
\eeq 
The projection of oscillators onto a given polarization restricts the phase space
so the $\hat q_0$ that enters the Cardy formula is smaller
than the units of momentum $q_0=N_3$ that correspond to $P$,
\beq
\hat q_0=q_0-q_J=q_0-\frac{6J^2}{c_{UV}}\,.
\eeq
This reproduces the entropy $S=2\pi\sqrt{N_1 N_2 N_3 -J^2}$ of the
supersymmetric rotating BMPV black hole and provides a
detailed account of its properties \cite{bmpv}.

To understand how black rings fit into this theory, observe that there
is another way in which angular momentum can be carried by the D1-D5
system. Each individual effective string has a fermionic ground state
that can be polarized to carry angular momentum $1$ (\ie $(1/2,1/2)$ of
the rotation group $SU(2)\times SU(2)\sim SO(4)$). In the untwisted
ground state there are $N_1 N_2$ such short strings, which can therefore
carry angular momentum
\beq
J_\psi=N_1 N_2\,,\qquad J_\phi=0\,.
\eeq
In this case angular momentum is present even in the absence of momentum
excitations. This ground state, which is a unique microstate,
corresponds to a class of systems generically known as {\it supertubes} \cite{MT},
since their spacetime realization is typically in terms of tubular
configurations of branes, in the present case a tube made of a single
Kaluza-Klein monopole, $n_3=1$. If there are several such monopoles,
$n_3>1$, then we are in a sector with $N_1 N_2/n_3=c_{UV}/6n_3$
strings of length $n_3$ and the angular momentum of the supertube is 
\beq\label{jsupert}
J_\psi=\frac{c_{UV}}{6n_3}.
\eeq

Supertubes have the right topology to be identified as constituents of
black rings and therefore provide string theory with the structure
required to accommodate different black objects with the same conserved
charges \cite{EE}. On the other hand, in order to obtain a macroscopic
degeneracy it seems necessary to have a `long-string' which
contains a thermal ensemble of thinly-spaced (small gap) momentum
excitations in much the same way as the BMPV black hole. Thus it is
natural to propose that the CFT of three-charge black rings decomposes
into two sectors, twisted and untwisted, with central charges $c_1$,
$c_2$ adding to $c_{UV}=c_1+c_2=6N_1N_2$ \cite{BK2}. The twisted ``BMPV
string'' carries the momentum charge and hence provides the entropy and
the $J_\phi$ component of the angular momentum. The untwisted ``supertube
string'' accounts for the anti-self-dual component of the angular
momentum, $J_\psi-J_\phi$, and is responsible for the tubular structure of the
configuration. Then the natural choice for the
central charge of the supertube string is, from \reef{jsupert}, $c_2=
6n_3 (J_\psi -J_\phi)$. 

This is an appealing, simple picture. If we take the BMPV string to be
maximally rotating, $q_J=N_3$, then its angular momentum is
\beq
J_\phi=\sqrt{\frac{c_{1}}{6} N_3}=
\sqrt{\left[N_1 N_2-n_3(J_\psi-J_\phi)\right]N_3}\,,
\eeq
and the entropy vanishes (to leading order). Some zero-area three-charge
black rings (those with ${\cal N}_1 {\cal N}_2=n_3 (J_\psi -J_\phi)$, so
$c_2=6{\cal N}_1 {\cal N}_2$) appear to be accurately described by these
formulae. However, the picture becomes problematic for configurations
with non-zero entropy. One apparent difficulty is that in this
description the transition from the black ring to the BMPV black hole is
smooth ---one simply eliminates the supertube sector, which does not
contribute any entropy--- in contradiction with the finite jump in the
area of the corresponding supergravity solutions. Indeed, when the
rotation of the BMPV string is less than maximal and there is a
degeneracy of states, the choice of $c_2$ for the supertube string
leaves too much central charge for the BMPV string and the microscopic
entropy is too large. This can be adjusted {\it ad hoc} but it is
unclear how to justify the larger value of $c_2$ that is needed to
reproduce the entropy of black rings \cite{BK2}.

\section{Other related developments}
\label{sec:otherstuff}

\subsection{Non-supersymmetric black rings, and the most general black
ring solution}

The most general solution constructed so far for non-supersymmetric
black rings is a seven-parameter family of non-supersymmetric black
rings which have three conserved charges, three dipole charges, two
unequal angular momenta, and finite energy above the BPS bound
\cite{EEF}. They have been found by solution-generating techniques
(boosts and U-dualities) applied to the five-dimensional dipole black
ring of \cite{RE} (see also \cite{HE,EE}). They are needed in order to
understand the thermal excitations of two- and three-charge supertubes. 

The supersymmetric limit of these solutions can only reproduce a
supersymmetric ring with three charges and at most two dipoles. A larger
family of non-supersymmetric black rings with nine-parameters
$(M,J_\psi,J_\phi,Q_{1,2,3},q_{1,2,3})$ is expected to exist, such that
the general solutions of \cite{EEMR1,BW,EEMR2,JGJG2} are recovered in
the supersymmetric limit. The nine parameters would yield three-fold
continuous non-uniqueness, the same as in the dipole rings of \cite{RE},
furnished by the non-conserved dipole charges $q_{1,2,3}$. 

In the limit in which the charges and dipoles vanish, this 9-parameter
solution would reduce to a 3-parameter vacuum solution generalizing that
of \cite{ER}. The third parameter would be angular momentum $J_\phi$ on
the $S^2$: this would be a ``doubly spinning" vacuum black ring. The most
promising method of obtaining this vacuum solution directly appears to
be the solution-generating techniques of \cite{5dsoliton}. However,
these methods seem to involve considerably more guesswork than the
relatively straightforward construction of the Kerr solution using
analogous methods in four dimensions.

It has been argued that the most general black ring solution should be
considerably larger than the 9-parameter family just discussed
\cite{larsen}. Toroidal compactification of eleven dimensional
supergravity to five dimensions yields $N=4$ supergravity, with 27
vectors and hence 27 independent electric charges. U-duality can be used
to eliminate 24 of these charges, so there is no loss of generality in
considering black holes carrying just 3 charges \cite{cvetichull},
corresponding to our $Q_i$. However, a general black ring should also
carry 27 dipoles. Using U-duality, the best that can be done in general
is to map this to a generating solution with 3 charges and 15 dipoles.
This indicates that the most general supersymmetric black ring solution
should have 19 parameters, and the most general non-supersymmetric black
ring will have 21 parameters. Constructing even the supersymmetric
solution will require some new ideas since the general form of
supersymmetric solutions of $N=4$ supergravity is a lot more complicated
than that of the $N=1$ $U(1)^3$ theory discussed above.

\subsection{`Small' black rings}

Supersymmetric rings with only two charges, and hence one dipole, have a
naked singularity instead of a horizon. However, the microscopic theory
still assigns them a finite entropy\footnote{When the internal space is
$K3\times S^1$. For $T^5$ the numerical prefactor is different and there
is a mismatch with the gravitational entropy.}
\beq\label{smallbr}
S=4\pi\sqrt{N_1 N_2 - n_3 J}\,.
\eeq
This is the degeneracy of fluctuations around the circular shape of a
supertube with less than maximal angular momentum, $J<N_1N_2/n_3$. In
the terms used in the previous section, it can be described as a
Bose-Einstein condensate of $J$ short strings of length $n_3$ (the
supertube), which account for the angular momentum, plus a thermal
ensemble of strings with degeneracy \reef{smallbr}
\cite{iishi,dabhetal}. 

Following the calculations performed for `small black holes'
\cite{dabh}, it has been possible to work out the regularization of the
singularity by higher-derivative corrections to the low energy effective
action in string theory. This requires the use of techniques developed
for four-dimensional $N=4$ supergravity, so the ring is compactified to
four dimensions by putting it on a Taub-NUT space. The
Bekenstein-Hawking-Wald entropy of the corrected horizon can then be
shown to reproduce \reef{smallbr} \cite{dabhetal}. 

A simple model for the dynamical appearance of the Bose-Einstein
condensate has been given in \cite{deboer}, where it is proposed that
the creation operators for short strings of length $n_3$ be assigned a
dipole $1/n_3$. The one-point functions of operators in the CFT dual to
the small black ring are non-trivial and match well with the above
description.

\subsection{Non-singular microstates and foaming black rings}

A main line of research that has provided a parallel motivation for much
of the work on black rings and related solutions is the ``fuzzball"
proposal for the fundamental structure of black holes \cite{fuzzball}.
According to this proposal, there should be some U-duality frame in
which black hole microstates admit a geometric description in terms of
non-singular, horizon-free supergravity solutions. The black hole is to be
regarded as an effective geometry describing an ensemble of microstates.

This programme has inspired the search for gravitating ``microstate"
solutions which are horizon-free and non-singular and which have the
same charges as a given black hole. In five dimensions, the complete
class of such (supersymmetric) solutions for the two-charge case is
known \cite{LMM} and one can indeed associate a solution with each
microstate of the underlying CFT. In this 2-charge case, the
system does not possess enough entropy to give rise to a macroscopic
horizon so 2-charge black holes do not exist (although higher-derivative
corrections can give rise to ``small black holes"). Nevertheless, the
microstate solutions do appear surprisingly ``black-hole-like", in
particular they exhibit a ``throat" region which closes off at a radius
set by the charges. If one computes the Bekenstein-Hawking entropy
associated with this radius then it agrees (up to a factor) with the
entropy obtained from CFT. 

Some 3-charge microstate solutions are known \cite{3charge} although too
few to make a non-vanishing contribution to the entropy. Progress
towards the construction of 3-charge solutions describing black ring
microstates has been made by adding a small amount of momentum $P$ as a
perturbation to 2-charge supertube solutions \cite{3chargepert}.

A large class of 3-charge solutions proposed as black hole and ring
microstate solutions has been constructed and studied in
refs.~\cite{ringbubble0,ringbubble}. Refs.~\cite{ringbubble} apply the
techniques described in subsec.~\ref{subsec:multi} to multi-center
Gibbons-Hawking base spaces with poles of positive and negative
residues. The solutions exhibit a rich topological structure related to
the ``resolution" of dipole sources into fluxes along new internal
cycles. However, in contrast to the solutions of the previous paragraph,
in these cases a mapping between these supergravity solutions and dual
CFT states has not been identified.

\section{Outlook} 

\subsection{Stability}

Thus far we have left entirely aside the important issue of the
classical dynamical stability of black rings. Supersymmetry should
ensure that supersymmetric black rings are stable to quadratic
fluctuations. Presumably, near-supersymmetric black rings are also
stable within a range of parameters. 

Vacuum black rings present a much more difficult case, since the study
of their linearized perturbations appears to be much harder than in the
case of Myers-Perry black holes:\footnote{
Even for these, the study of gravitational perturbations may only be
tractable analytically in special cases with enhanced symmetry arising
when some angular momenta coincide \cite{KLR}.} even the
equation for massless scalar fields in the presence of the black ring
does not appear to be separable.
Therefore, the study of gravitational perturbations
around such backgrounds might require evolving the equations of motion
for such perturbations numerically. This does not sound conceptually
challenging so it will be interesting to see whether progress can be
made this way.

For the time being, some qualitative and semi-quantitative arguments
have been advanced. The black ring at the cusp between the thin and fat
black ring branches must certainly be unstable: by throwing at it any
small amount of matter that adds mass but not angular momentum, there is
no other black ring that the system can evolve into and therefore it
must backreact violently \cite{GRG}. Qualitatively, it is expected that thin black
rings suffer from the Gregory-Laflamme instability \cite{GL}, which would grow
lumps on the ring and whose evolution is at present uncertain
\cite{ER,HoMy,toappear}. 

A study of the topology of the phase diagram of black rings and MP black
holes indicates that, precisely at the cusp, at least one unstable mode
is added when going from thin to fat black rings, \ie the latter
should be more unstable \cite{LoTeAr}. This is consistent with an analysis of
radial perturbations \cite{toappear}, against which all fat black rings appear to
be unstable, while thin black rings are radially stable. This suggest
that the vacuum black rings with a single spin described in this paper
can be stable only if the GL instability switches off for thin black
rings with small enough $j$, a possibility that deserves further study.
Vacuum black rings with two spins (yet to be constructed) presumably
suffer also from superradiant instabilities peculiar to black objects
with a rotating two-sphere \cite{odias}. The dipole charge in black rings may
help stabilize them against some perturbations, like GL modes, in
particular near the extremal limit.

\subsection{Generalizations}

The number of degrees of freedom of gravity increases with the spacetime
dimension $D$, so it is natural to expect more complex dynamics as $D$
grows. In $D<4$ the dynamics is so constrained that gravity has no
propagating degrees of freedom. In $D=4$ gravity does propagate, but it
is still highly constrained, as the black hole uniqueness theorems
illustrate. The discovery of black rings shows that $D=5$ allows for
more freedom but the dynamics is still amenable to detailed study.
Gravity in $D>5$ remains largely unexplored, but there are
indications that black holes (even spherical ones) possess qualitatively
new features \cite{EM}. 

It is natural to wonder whether black rings of horizon topology
$S^1\times S^{D-3}$ exist for $D>5$. What about other topologies in
$D>5$, \eg $S^1\times S^1 \times S^2$, $S^3\times S^3$ etc? These
possibilities are all consistent with the higher-dimensional topology
theorem of \cite{topology}.\footnote{It will be up to their discoverers
to find a good name for these black holes.} 

Heuristically, the balance of forces in thin black rings happens between
centrifugal repulsion and tension \cite{HoMy,toappear}, which are
independent of the number of dimensions since both forces are confined
to the plane of the ring. The dimension-dependent gravitational force
decays faster with the distance, so it plays a role only in the
equilibrium of rings at small radii. This suggests that thin black
rings should also exist in $D>5$, and, like in $D=5$, be unstable against
radial perturbations. Observe that, since there is no bound on the
angular momentum of MP black holes with a single spin in $D>5$
\cite{MP}, the existence of these rotating black rings {\it would}
automatically imply the violation of black hole uniqueness.

Another possibility is that there might exist black rings with less
symmetry than any known solution. It has been proved that a
higher-dimensional stationary rotating black hole must admit a
rotational symmetry \cite{HIW}. However, known black rings (indeed, all
known $D \ge 5$ black hole solutions) have multiple rotational
symmetries. This has led to the suggestion that there may exist black
hole solutions with less symmetry than the known solutions \cite{harv}.
An example would be a vacuum black ring solution with the same charges
as the one of \cite{ER} but lacking the ``accidental" rotational
symmetry $\partial/\partial \phi$ on the $S^2$.

\subsection{From microscopics to macroscopics}

Typically, the approach to black hole entropy calculations has been to
start from a black hole solution, obtain its entropy from the area of
the event horizon, and then try to reproduce this result statistically
from a microscopic theory. However, one could just as well work
backwards by constructing a microscopic model that gives rise to a
macroscopic entropy and thereby predicting the existence of an
associated black hole solution. For example, a formula for the
microscopic entropy of the yet-to-be-found 9-parameter black ring
solution discussed above was proposed in \cite{larsen} based on
U-duality of the IR theory\footnote{Like in section~\ref{sec:micro}, the
validity of this formula depends on the absence of finite ring radius
corrections, which is not guaranteed.}. It will be interesting to see
whether this kind of approach can be pushed further, for example to
predict properties of new black holes in $D>5$. In general, such
predictions are not much help in finding black hole solutions. However,
special cases in which some supersymmetry is preserved may be more
tractable, as we have seen is the case in $D=5$.

\section*{Acknowledgments} We are grateful to our collaborators on this
topic, Henriette Elvang, Pau Figueras, Gary Horowitz, and David Mateos.
Although Andrew Chamblin never authored a paper on black rings, he
played a crucial role at different stages in the early developments of
the subject. RE is supported in part by DURSI 2005 SGR 00082, FPA
2004-04582-C02-02 and EC FP6 program MRTN-CT-2004-005104. HSR is a Royal Society University Research Fellow.

%
%
%

\end{document}